\documentclass[pdflatex,sn]{sn-jnl}

\usepackage{graphicx}%
\usepackage{multirow}%
\usepackage{amsmath,amssymb,amsfonts}%
\usepackage{amsthm}%
\usepackage{mathrsfs}%
\usepackage[title]{appendix}%
\usepackage{xcolor}%
\usepackage{textcomp}%
\usepackage{manyfoot}%
\usepackage{booktabs}%
\usepackage{algorithm}%
\usepackage{algorithmicx}%
\usepackage{algpseudocode}%
\usepackage{listings}%

\usepackage{lineno}
\theoremstyle{thmstyleone}%
\theoremstyle{thmstyletwo}%

\theoremstyle{thmstylethree}%

\begin{document}

\title[Article Title]{MolCluster: Integrating Graph Neural Network with Community Detection for Coarse-Grained Mapping} 

\author[1,2]{\fnm{Zhixuan Zhong}} 

\author[3]{\fnm{Linbo Ma}} 

\author*[1,2]{\fnm{Jian Jiang}}\email{jiangj@iccas.ac.cn}

\affil*[1]{\orgdiv{Beijing National Laboratory for Molecular Sciences, State Key Laboratory of Polymer Physics and Chemistry, Institute of Chemistry}, \orgname{Chinese Academy of Sciences}, \orgaddress{\city{Beijing}, \postcode{100190}, \country{P. R. China}}}

\affil*[2]{\orgname{University of Chinese Academy of Sciences}, \orgaddress{\city{Beijing}, \postcode{100049}, \country{P. R. China}}}

\affil*[3]{\orgname{China Academy of Safety Science and Technology}, \orgaddress{\city{Beijing}, \postcode{100012}, \country{P. R. China}}}


\abstract{Coarse-grained (CG) modeling simplifies molecular systems by mapping groups of atoms into representative units. However, traditional CG approaches rely on fixed mapping rules, which limit their ability to handle diverse chemical systems and require extensive manual intervention. Thus, supervised learning-based CG methods have been proposed, enabling more automated and adaptable mapping. Nevertheless, these methods suffer from limited labeled datasets and the inability to control mapping resolution, which is essential for multiscale modeling. To overcome these limitations, we propose MolCluster, an unsupervised model that integrates a graph neural network and a community detection algorithm to extract CG representations. Additionally, a predefined group pair loss ensures the preservation of target groups, and a bisection strategy enables precise, customizable resolution across different molecular systems. In the case of the downstream task, evaluations on the MARTINI2 dataset demonstrate that MolCluster, benefiting from its label-free pretraining strategy, outperforms both traditional clustering and supervised models. Overall, these results highlight the potential of MolCluster as a core model for customizable and chemically consistent CG mapping.}

\keywords{Coarse-Grained Mapping, Graph Neural Networks, Community Detection, Pretraining}



\maketitle

\section*{Introduction}\label{sec1}

Coarse-grained (CG) mapping is the primary step in CG modeling, aimed at extracting the key substructures from atomic-level molecular models to create simplified CG bead models \cite{Noid2022, Jin2022, Mancardi2023, Kidder2024}. This approach significantly improves the computational efficiency of molecular dynamics simulations by reducing degrees of freedom, enabling large-scale and long-time simulations. Moreover, the chemically grounded CG representations also providers more refined and meaningful feature representations for machine learning methods. \cite{Yue2024, Kramer2023} Achieving accurate mapping is critical in CG modeling, where the key challenge lies in preserving maximal atomic-level information while simultaneously capturing the essential molecular structures.\cite{Noid2013, Saunders2013, Jin2022, Khot2019} However, traditional CG mapping methods usually rely on manual intervention and expert strategies, such as functional group,\cite{Rappoport2009, Ertl2017, Mukherjee2023, Colmenarejo2025} rigid and flexible region, \cite{Ricci2019, Whiteley} macromolecular monomer,\cite{Zhu2022, Dhamankar2021, Shi2023, Borges-Araujo2023, Ingolfsson2014, Wang2025} dynamic grouping \cite{Yang2023, Foley2020, Kidder2024}, and force field predefined rules. \cite{Marrink2007, Souza2021} These strategies are manually designed and lack flexibility, thereby limits their ability to adapt to the complexity and diversity of molecular systems.

To address this issue, automated CG mapping methods based on graph neural networks (GNNs) have become a growing research interest in recent years \cite{Li2020, Errica2021, Martino2024, Zhong2025}. Compared to traditional molecular graph clustering, GNNs exhibit superior expressive power by simultaneously capturing bond orders, atomic types, charges, electronegativity, and so on.\cite{Du2024, Wu2023-2, Deng2023, Li2022, Hu2024, Hu2025}. Errica and coworkers combined GNNs with the Wang–Landau sampling algorithm to efficiently estimate mapping entropy, reducing computational cost and enabling broader exploration of mapping spaces \cite{Errica2021}. While this method showed promising results in protein systems, its transferability to diverse molecular structures remains uncertain and may require large-scale training across varied datasets. In our previous work, \cite{Zhong2025} the DSGPM-TP model was proposed to treat  the CG mapping process as a dual-task problem, which includes atomic clustering (a graph partitioning task) and CG bead type prediction (a classification task). By integrating spectral clustering with GNNs, DSGPM-TP offers an alternative to manually designed CG mapping rules, providing a more data-driven approach. However, despite its promising framework, the DSGPM-TP model still exhibits several limitations in practical applications. Firstly, the mapping process in DSGPM-TP still heavily relies on predefined rules of force field, such as the four-to-one mapping strategy in the MARTINI model. \cite{Marrink2007, Souza2021} Secondly, DSGPM-TP requires extracting CG mapping data from literature sources, where datasets are often small, incomplete, and lack diversity, further constraining the model's generalization ability across various systems. Large and high-quality labeled datasets are also a prerequisite for many deep learning-based CG tasks \cite{Ricci2023, Wang2019, Sahrmann2025, Unke2021}, suggesting that unsupervised methods such as variational autoencoders may offer an effective solution.\cite{Wang2019-2} Finally, DSGPM-TP requires a given number of CG beads and cannot achieve adaptive or controllable CG resolution. These limitations indeed hinder its practical applicability in multiscale molecular simulations. 

To overcome these challenges, in this work, we propose a novel model, MolCluster, which integrates GNNs with Leiden community detection algorithm\cite{Traag2019}.  The Leiden algorithm, which optimizes network modularity, enables the automatic identification of CG groups without requiring the setting of the number of beads and their partitioning scheme. This design allows MolCluster to perform effectively in few-shot learning settings and reduces its dependence on predefined CG bead counts. Moreover, this model supports downstream tasks such as the construction of mapping schemes with user-defined resolution and CG group type prediction. Consequently, MolCluster offers a more flexible and efficient approach to significantly advancing CG modeling by improving adaptability, structural fidelity, and applicability across diverse molecular systems.

\section*{Results}\label{sec2}

\begin{figure}[htbp]
	\centering
	\includegraphics[scale=0.2]{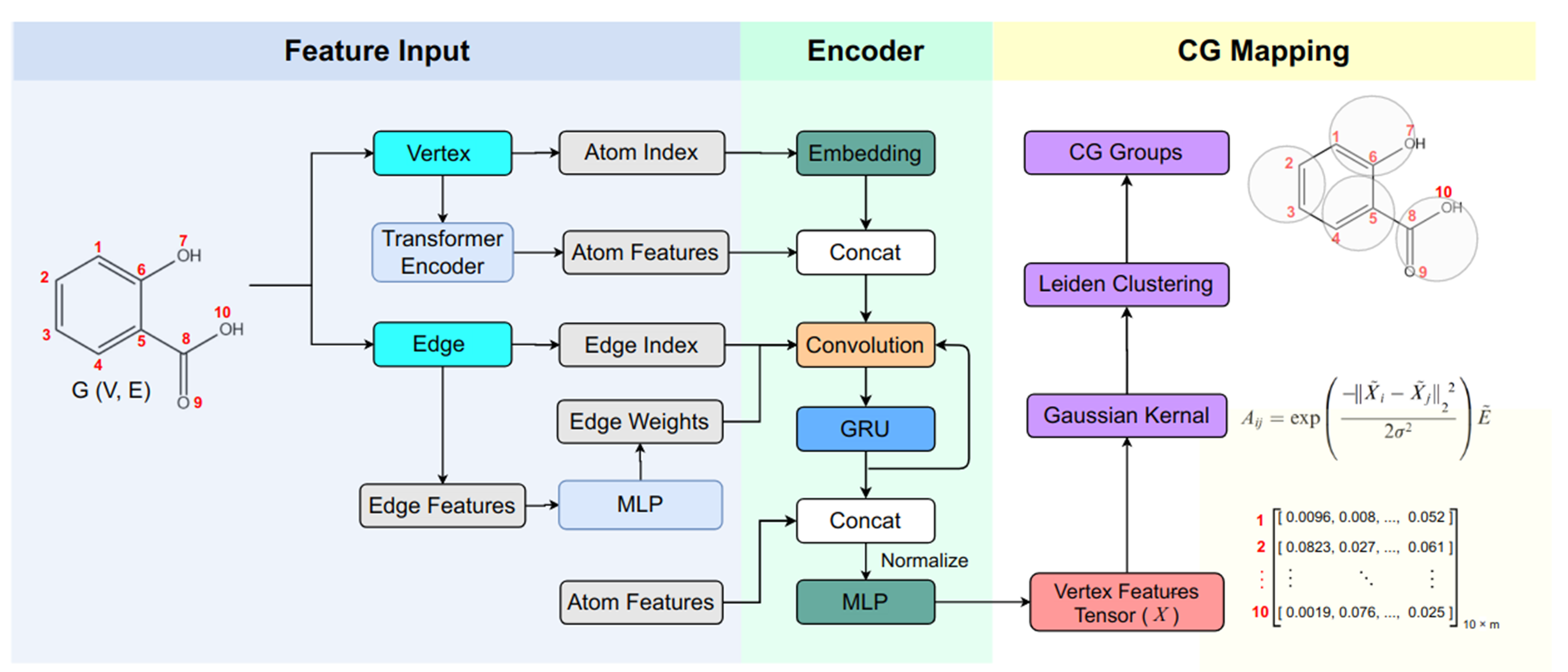}
	\caption{Architecture of MolCluster.}
	\label{fig1}
\end{figure}

As a graph-based clustering model, MolCluster uses GNN architecture with a Transformer encoder \cite{Vaswani2017} to effectively extracts both local chemical features and global molecular representations, as illustrated in Figure \ref{fig1}.  Training is guided by the loss function(Eq. \ref{lossfunction}) that combines  of triplet loss and predefined group pair loss to ensure structural consistency and group preservation. A comprehensive demonstration of our MolCluster model is provided in the “Method” section below. Beyond traditional CG mapping, MolCluster supports user-defined group preservation and adjustable CG resolution, and it can also serve as a pretraining model for supervised CG mapping approaches. 


\subsection*{User-Defined Group Preservation via Predefined Group Pair Loss}

\begin{figure}[htbp]
	\centering
	\includegraphics[scale=0.3]{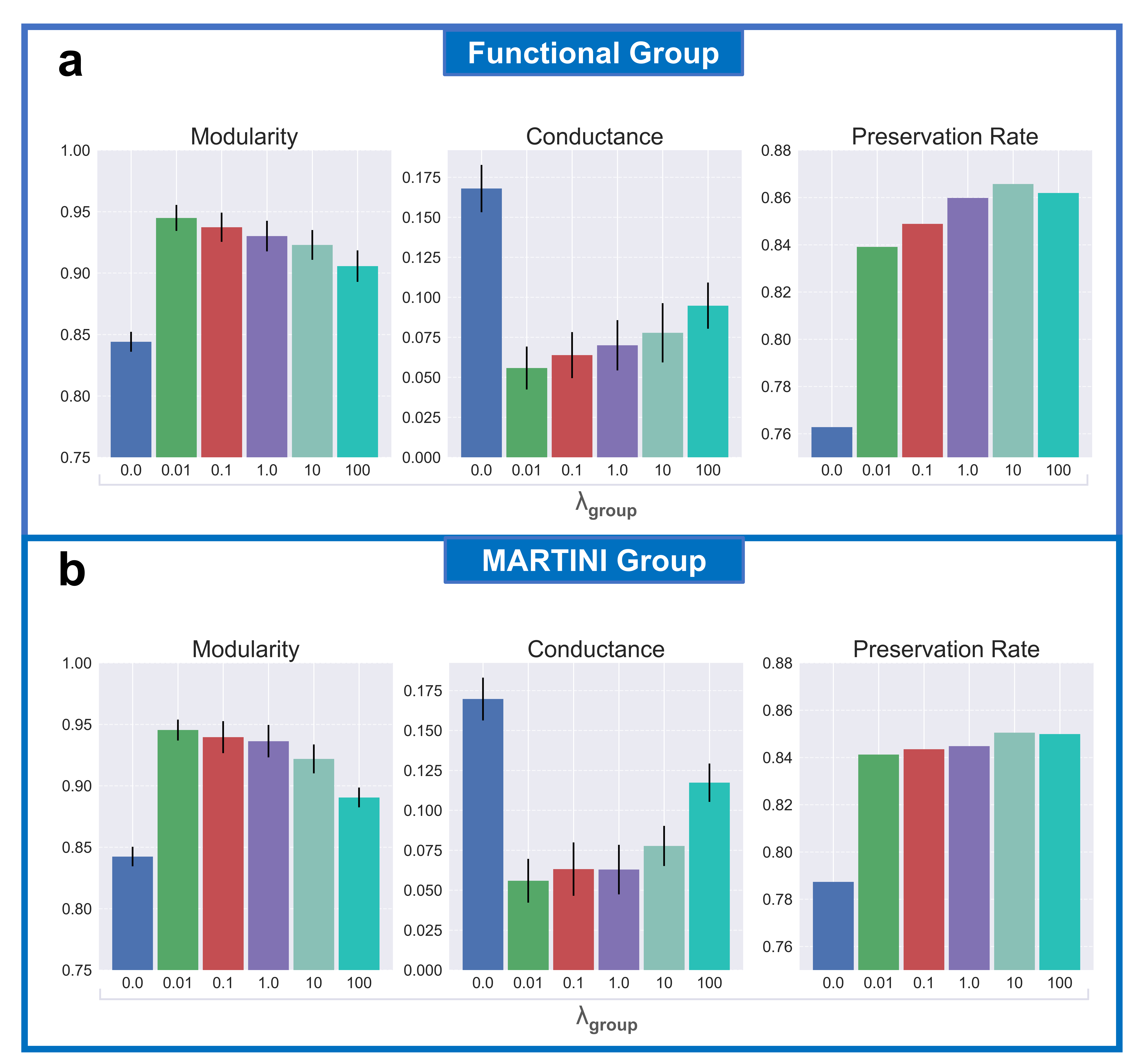}
	\caption{Modularity, conductance, and preservation rate in the case of predefined group of (a) functional group and (b) MARTINI group.  Here, higher modularity (better community partitioning) and preservation rate (more retained target groups) are preferred; lower conductance (sparser inter-group connections) is better.}
	\label{fig2}
\end{figure}

To enhance the preservation of meaningful substructures, a predefined group pair loss (Eq.\ref{lossfunction}) is introduced during training, enabling controllable CG mapping guided by user-defined group integrity. Specifically, 38 common functional groups (\textit{Functional\_Group\_SMARTS.txt} in SI) are predefined according to SMARTS patterns \cite{Daylight_SMARTS} using RDKit. These groups serve as chemical constraints to ensure structural completeness during mapping. As shown in Figure \ref{fig2}a, the group preservation rate (defined as the proportion of predefined groups that are fully preserved in MolCluster’s clustering results) is approximately 0.76 without incorporating the predefined group pair loss. This indicates that the method is inherently capable of capturing essential chemical substructures to some extent. When the group pair loss is applied, the preservation rate increases significantly, confirming that the model can better preserve the integrity of user-defined groups, producing CG mappings that are more consistent with expert knowledge.

Meanwhile, increasing the weight of group pair loss ($\lambda_{\text{group}}$) leads to a slight decrease in modularity but substantially give a rise in  conductance (see SI for more details), suggesting that excessive weighting may overemphasize chemical constraints at the expense of global topological consistency. Therefore, we set $\lambda_{\text{group}}=1.0$ to balance functional group preservation and overall structural coherence. Additionally, MolCluster is also evaluated on 38 MARTINI-specific groups (\textit{MARTINI\_Group\_SMARTS.txt} in SI) which were extracted from 722 MARTINI mapping cases \cite{Zhong2025} for further testing of the preservation rate. As shown in Figure \ref{fig2}b, MolCluster preserves MARTINI mapping groups effectively during CG partitioning as well. In brief, these results demonstrate that MolCluster can flexibly incorporate chemical knowledge and predefined structural constraints to achieve chemically consistent and expert mapping schemes.

\subsection*{Customizable CG Resolution via Bisection Strategy}

The ability to customize CG resolution is essential for multiscale molecular modeling, where different levels of CG mapping are required to balance computational efficiency and structural fidelity. While predefined group preservation ensures the chemical integrity of each predefined group,, it does not directly control the resolution of the CG representation. To enable controllable CG resolution across diverse molecular systems, we introduce a parameter $r$ to define the CG resolution:
\begin{equation}
	r = \frac{N_{\text{groups}}}{N_{\text{atoms}}},
\end{equation}
where $N_{\text{groups}}$ is the number of CG beads after mapping, and $N_{\text{atoms}}$ is the total number of atoms in the original molecule. A smaller value of $r$ corresponds to a coarser mapping, while a larger $r$ indicates a finer representation. In practice, $r$ can be adjust by regulating $N_{\text{groups}}$. 

\begin{figure}[htbp]
	\centering
	\includegraphics[scale=0.3]{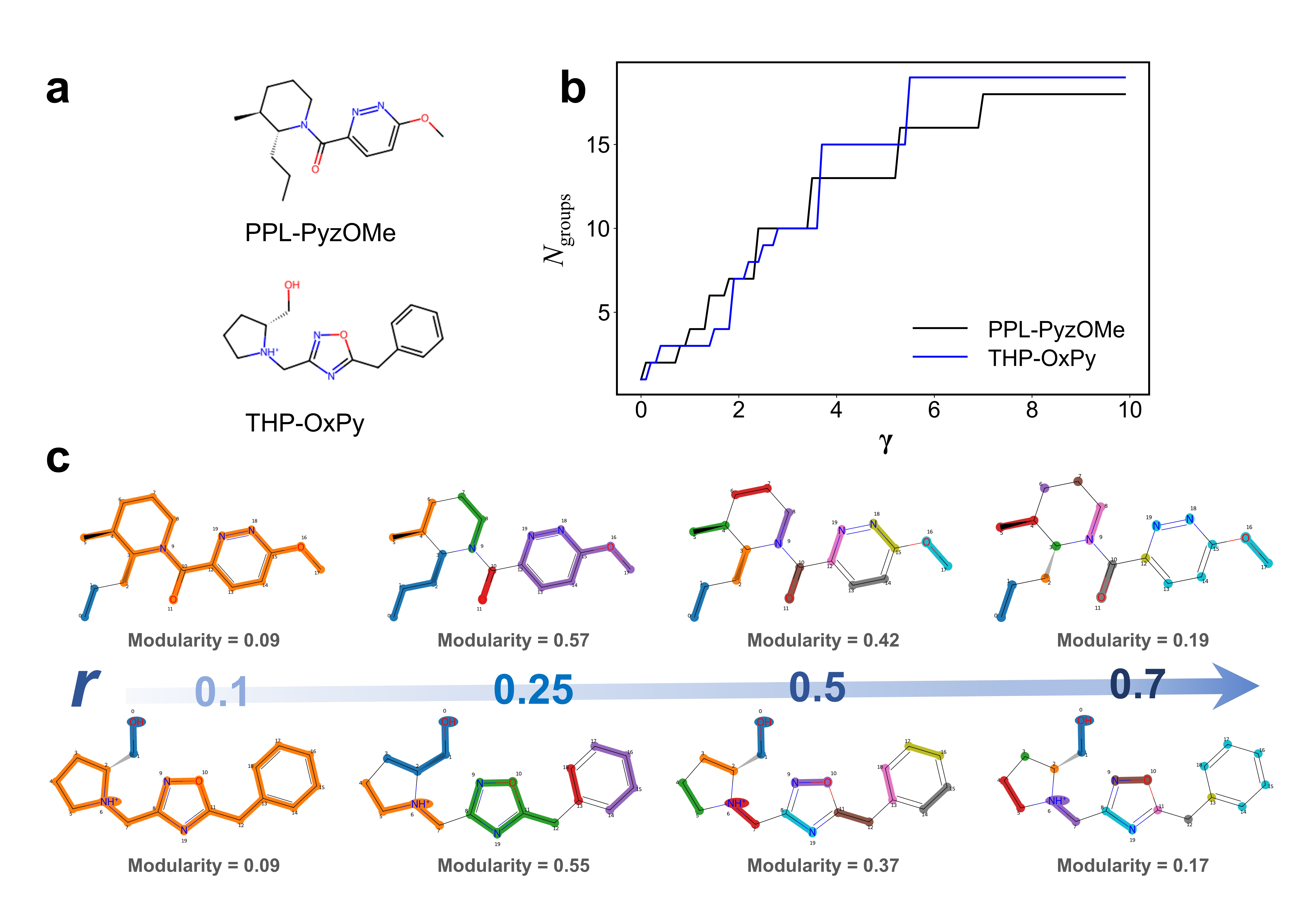}
	\caption{(a) Molecular structure; (b) relationship between number of CG groups ($N_{\text{groups}}$) and resolution parameter $\gamma$; (c) CG Mappings of molecules with different CG resolution ($r$).}
	\label{fig3}
\end{figure}

Since the number of CG groups $N_{\text{groups}}$ increases monotonically with the resolution parameter $\gamma$ in Eq.\ref{modularity} (Figure S1), it is possible to determine an optimal $\gamma$ that yields a target CG resolution $r_{\text{target}}$. For example, in the cases of molecules PPL-PyzOMe and THP-OxPy (Figure \ref{fig3}a), the number of CG groups consistently increases with $\gamma$, as shown in Figure \ref{fig3}b, confirming this monotonic behavior. Therefore, an adaptive parameter tuning strategy based on the bisection method is introduced to optimize $\gamma$, ensuring that $N_{\text{groups}}$ converges to a target value $N_{\text{target}}$ corresponding to the desired resolution $r_{\text{target}}$. The core idea is to iteratively narrow the search interval for $\gamma$ until $N_{\text{groups}}$ is sufficiently close to $N_{\text{target}}$. The pseudocode (more details can be found in SI)) is illustrated in Algorithm \ref{alg:optimize_gamma}.

\begin{algorithm}
	\caption{Bisection Optimization of $\gamma$ for Target CG Resolution}
	\label{alg:optimize_gamma}
	\begin{algorithmic}[1]
		\Require Target group number $N_{\text{target}}$, initial range $\gamma_{\text{min}}$, $\gamma_{\text{max}}$, tolerance $\epsilon$
		\Ensure Optimized $\gamma$
		\While{$\gamma_{\text{max}} - \gamma_{\text{min}} > \epsilon$}
		\State $\gamma = (\gamma_{\text{min}} + \gamma_{\text{max}})/2$
		\State Run MolCluster and compute $N_{\text{groups}}$
		\If{$|N_{\text{groups}} - N_{\text{target}}| < \epsilon$}
		\State \Return $\gamma$ \Comment{Optimal parameter found}
		\ElsIf{$N_{\text{groups}} > N_{\text{target}}$}
		\State $\gamma_{\text{max}} = \gamma$ \Comment{Decrease $\gamma$ to merge groups}
		\Else
		\State $\gamma_{\text{min}} = \gamma$ \Comment{Increase $\gamma$ to split groups}
		\EndIf
		\EndWhile
		\State \Return $\gamma$ \Comment{Return optimized parameter}
	\end{algorithmic}
\end{algorithm}

Compared to a fixed-$\gamma$ approach, this self-adaptive method provides significantly higher robustness, ensuring consistent CG resolution across molecular systems with varying sizes, structures, and connectivity patterns. This method was validated on the PPL-PyzOMe and THP-OxPy molecular systems. As shown in Figure \ref{fig3}c,  the bisection-based optimization consistently converges to the target resolution, providing nearly identical CG levels across different systems. Beyond small-molecule systems, this approach is extensible to complex structures such as polymers and proteins, where resolution control is critical for multiscale modeling. By combining the adjustability of resolution and the preservation of desired fragments (enabled by predefined group pair loss), MolCluster makes complex multi-scale modeling feasible.

\subsection*{MolCluster as a Label-Free Pretraining Model for Supervised Methods}

In the previous section, a bisection-based adaptive resolution strategy is introduced, which relies on a tunable parameter $\gamma$ to ensure that different molecular systems achieve comparable CG resolution. Since MolCluster does not require labeled data to learn molecular representation, it can also serve as a label-free pretraining model to enhance the performance of supervised CG mapping approaches, such as DSGPM-TP\cite{Zhong2025}.

\begin{table}[!htbp]
	\centering
	\caption{Clustering performance of different models.}
	\label{martini_clustering_result}
	\footnotesize
	\setlength{\tabcolsep}{4pt}
	\renewcommand{\arraystretch}{1.2}
		\begin{tabular}{ccccc}
			\hline
			Method & AMI & Cut Prec. & Cut Recall & Cut F1-score \\
			\hline
			HDBSCAN \cite{Campello2013} & 0.3534$_{(0.0421)}$ & 0.4155$_{(0.0672)}$ & 0.2594$_{(0.0259)}$ & 0.2702$_{(0.0367)}$ \\
			FINCH \cite{Sarfraz2019} & 0.5348$_{(0.0514)}$ & 0.3915$_{(0.0385)}$ & 0.6499$_{(0.0572)}$ & 0.4684$_{(0.0468)}$ \\
			h-NNE \cite{Sarfraz2022} & 0.2648$_{(0.0642)}$ & 0.3116$_{(0.0241)}$ & 0.656$_{(0.0510)}$ & 0.4001$_{(0.0308)}$ \\
			Graclus \cite{Dhillon2007} & 0.4107$_{(0.0828)}$ & 0.2625$_{(0.0078)}$ & 0.6449$_{(0.0239)}$ & 0.3677$_{(0.0117)}$ \\
			METIS \cite{Karypis1998} & 0.7277$_{(0.0000)}$ & 0.5786$_{(0.0000)}$ & 0.6097$_{(0.0000)}$ & 0.5926$_{(0.0000)}$ \\
			Spectral Cluster \cite{Ng2001}  & 0.7538$_{(0.0607)}$ & 0.5408$_{(0.0123)}$ & 0.5323$_{(0.0123)}$ & 0.5362$_{(0.0123)}$ \\
			DSGPM \cite{Li2020} & 0.8328$_{(0.0294)}$ & 0.6848$_{(0.0530)}$ & 0.6807$_{(0.0527)}$ & 0.6825$_{(0.0528)}$ \\
			DSGPM-TP \cite{Zhong2025}  &  0.8493$_{(0.0290)}$ & 0.7159$_{(0.0587)}$ & 0.7125$_{(0.0594)}$ & 0.7139$_{(0.0590)}$ \\
			MolCluster(ours)  & 0.8818$_{(0.0084)}$ & 0.7859$_{(0.0150)}$ & 0.7853$_{(0.0172)}$ & 0.7854$_{(0.0161)}$   \\
			\textbf{MolCluster$_\text{fine\_tune}$(ours)} & \textbf{0.9107$_{(0.0046)}$} & \textbf{0.8304$_{(0.0103)}$} & \textbf{0.8345$_{(0.0127)}$} & \textbf{0.8321$_{(0.0113)}$}  \\
			\hline
		\end{tabular}
\end{table}

\begin{table}[!htbp]
	\centering
	\caption{Type prediction performance of different models.}
	\label{martini_type_result}
	\footnotesize
	\setlength{\tabcolsep}{4pt}
	\renewcommand{\arraystretch}{1.2}
		\begin{tabular}{cccc}
			\hline
			Method & Type Prec. & Type Recall & Type F1-score \\
			\hline
			DSGPM \cite{Li2020} & 0.0138$_{(0.0109)}$ & 0.0158$_{(0.0096)}$ & 0.0129 $_{(0.0120)}$ \\
			DSGPM-TP \cite{Zhong2025}  & 0.8501$_{(0.1161)}$ & 0.8100$_{(0.0995)}$ & 0.8177 $_{(0.1060)}$ \\
			MolCluster(ours)  & 0.8844$_{(0.0069)}$ & 0.8558$_{(0.0061)}$ & 0.8624$_{(0.0062)}$ \\
			\textbf{MolCluster$_\text{fine\_tune}$(ours)}  & \textbf{0.8931$_{(0.0089)}$} & \textbf{0.8596$_{(0.0057)}$} & \textbf{0.8672$_{(0.0057)}$} \\
			\hline
		\end{tabular}
\end{table}

To comprehensively evaluate MolCluster's performance and highlight its advantage as a pretraining model, extensive tests are conducted on the MARTINI2 mapping dataset.\cite{Zhong2025} MolCluster is compared against several representative clustering methods, including density-based HDBSCAN~\cite{Campello2013}, hierarchical clustering approaches FINCH~\cite{Sarfraz2019} and h-NNE~\cite{Sarfraz2022}, graph-partitioning algorithms Graclus~\cite{Dhillon2007} and METIS~\cite{Karypis1998}, and classical spectral clustering~\cite{Ng2001}. We also compared it to the supervised DSGPM-TP model, which directly performs atom grouping and type prediction on molecular graphs.\cite{Zhong2025}

\begin{figure}[htbp]
	\centering
	\includegraphics[scale=0.2]{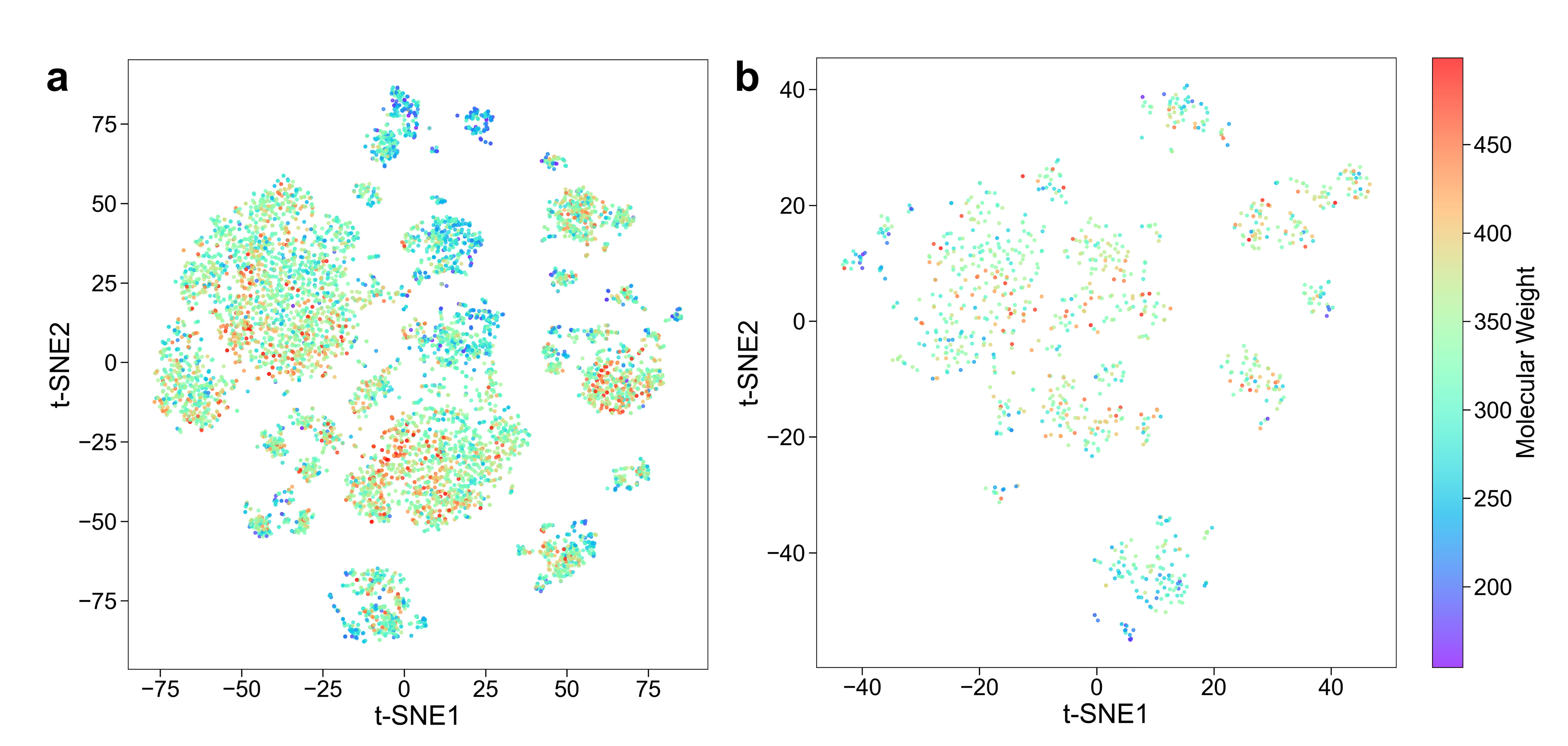}
	\caption{Visualization of molecular representations learned by MolCluster via t-SNE. Representations are extracted from the (a) training and (b) test set of the pretraining dataset. Each point is colored by its corresponding molecular weight.}
	\label{fig4}
\end{figure}

Table \ref{martini_clustering_result} demonstrates that MolCluster achieves superior clustering performance compared to both unsupervised clustering methods and the DSGPM model. In particular, MolCluster achieves state-of-the-art clustering precision and recall, indicating a more accurate delineation of CG group boundaries and a reduction in misassignments. With the finetuning stage, all performance metrics further are further improved, with the F1-score reaching 0.8321, highlighting enhanced generalization and predictive accuracy. In the CG bead type prediction task, as shown in Table \ref{martini_type_result}, MolCluster and its finetuned version also exhibit outstanding performance. Compared to DSGPM-TP, MolCluster represents a notable improvement, achieving a precision of 0.8931. This demonstrates that MolCluster not only provides accurate CG mappings but also predicts bead types with high fidelity. Furthermore, the t-SNE visualizations of the pretrained feature embeddings (Figure \ref{fig4}) provide additional evidence of the effectiveness of pretraining. In the training set (Figure \ref{fig4}a), molecules with similar molecular weights are organized into well-defined clusters, and a comparable pattern is observed in the test set (Figure \ref{fig4}b), demonstrating strong generalization capability. These results indicate that MolCluster learns chemically meaningful and transferable representations during pretraining, thereby enhancing its performance in downstream tasks such as forcefiled-rule mapping. In a word, with predefined group pair loss ensuring chemical interpretability and customizable CG resolution boosting adaptability throughout pretraining, MolCluster stands out as a reliable pretraining platform that integrates representational strength, structural integrity and system adaptability for CG modeling.

\section*{Discussion}\label{sec12}

In summary, this work proposes MolCluster, a model that integrating graph neural network and community detetion algorithm for automated molecular coarse-grained mapping. By incorporating a predefined group pair loss, the method preserves structural integrity and chemical interpretability during CG mapping, and a self-adaptive bisection strategy enables precise control of the CG resolution across diverse molecular systems. Furthermore, MolCluster can be served as a label-free pretraining model, providing transferable molecular embeddings that enhance supervised approaches on limited datasets. Evaluation on the MARTINI2 benchmark demonstrates that MolCluster achieves superior partitioning accuracy and bead-type prediction compared to traditional clustering methods and supervised baselines. These results highlight MolCluster's potential for generalizable and chemically consistent coarse-grained modeling of complex molecular systems. Future work will focus on extending the framework of MolCluster to handle larger molecular systems, incorporating dynamic information from molecular dynamics trajectories, and integrating force-field parameterization to achieve end-to-end automated coarse-grained modeling in AMOFMS\cite{Zhong2025}.




\section*{Methods}

\subsection*{Architecture of MolCluster}

MolCluster is a graph clustering model based on an improved GNN, as illustrated in Figure ~\ref{fig1}. It takes a molecular graph as input, where atoms and chemical bonds are represented as node and edge features, respectively. Atom types are first embedded into high-dimensional feature vectors and concatenated with representations encoded by a Transformer encoder. Edge features are processed by a multilayer perceptron (MLP) to generate dynamic edge weights. During multiple rounds of message passing and updates, node features aggregate information from neighboring nodes, effectively capturing local molecular structures. Based on the learned node features, a Gaussian kernel function is applied to compute edge weights, forming a weighted adjacency matrix. Finally, MolCluster employs the Leiden algorithm, \cite{Traag2019} an efficient community detection method that optimizes modularity, to identify community structures within the molecular graph and achieve CG mapping. The following sections will describe how the improved GNN generates a high-precision adjacency matrix and how the Leiden algorithm performs CG mapping based on this matrix.

\subsection*{Adjacency Matrix via an Improved GNN}

\subsubsection*{Initial Embedding}
Each atom is encoded as a one-hot vector $\text{X}$ according to its atomic number, which is then projected into a high-dimensional embedding space using a query matrix:
\begin{equation}
	\mathbf{X} = \text{Em}\mathbf{X},
\end{equation}
where the query matrix $\text{Em} \in \mathbb{R}^{N \times d}$, with $N$ denoting the number of atoms and $d$ the embedding dimension. The resulting embedding vectors are concatenated with atomic features encoded by a Transformer encoder \cite{Vaswani2017} :
\begin{equation}
	\mathbf{X}^{0} = \textbf{Concat}(\mathbf{X}, \mathbf{F_u}),
\end{equation}
where $\mathbf{F_u} \in \mathbb{R}^{N \times 32}$ includes atomic properties such as aromaticity, degree, and charge (Table~\ref{Chap_4_MolCluster:molcluster_feature_input}).

\begin{table}[htbp]
	\centering
	\renewcommand\arraystretch{1.0}
	\caption{Input features and their descriptions.}
	\label{Chap_4_MolCluster:molcluster_feature_input}
	\footnotesize
	\setlength{\tabcolsep}{4pt} 
	\renewcommand{\arraystretch}{1.2} 
		\begin{tabular}{ccccc}
			\hline
			Type & \textbf{Features} & \textbf{Description}  & \textbf{Form} & \textbf{Size}\\
			\hline
			\multirow{12}{*}{Node (Atom)} 
			& Atomic number & - & Embedding & 30\\
			& Aromaticity & Part of an aromatic system. & One-Hot & 1 \\
			& Degree & Degree of the atom in the molecule. & One-Hot & 9 \\
			& Number of H & Number of bonded hydrogen atoms. & One-Hot & 9 \\
			& Hybridization & SP, SP2, SP3, SP3D, SP3D2, UNSPECIFIED. & One-Hot & 6 \\
			& Charge & Gasteiger charge. & Float & 1 \\
			& Mass & Atomic mass. & Float & 1 \\
			& Radius & Atomic radius. & Float & 1 \\
			& Electron affinity & - & Float & 1 \\
			& First ionization energy & - & Float & 1 \\
			& Electronegativity & - & Float & 1 \\
			& Valence electrons & - & Float & 1 \\
			\hline
			\multirow{4}{*}{Edge (Bond)} 
			& Bond Directionality & None, Beginwedge, Begindash, etc. & One-Hot & 7 \\
			& Bond Type & Single, Double, Triple, or Aromatic. & One-Hot & 4 \\
			& Bond Length & - & Float & 1 \\
			& Bonded Potential & $E_{\mathrm{bonded}} = \frac{1}{2}k(r-r_0)^2$ & Float & 1\\
			\hline
	\end{tabular}
\end{table}

\subsubsection*{Edge Feature Transformation}
The edge features $\mathbf{E}_{uv}$ are processed through an MLP to generate a weight matrix relevant to edge conditions:
\begin{equation}
	\phi^e(\mathbf{E}_{uv}) \in \mathbb{R}^{d \times d},
\end{equation}
where $\mathbf{E}_{uv} \in \mathbb{R}^{F_e}$ and $F_e$ denotes the number of edge features. During each iteration $t$, edge features $\mathbf{E}_{uv}$ are recalculated to integrate updated node features $\mathbf{X}_u^{t}$ and $\mathbf{X}_v^{t}$. The edge weight $w_{uv}$ between nodes $u$ and $v$ is computed using a Gaussian kernel:
\begin{equation}
	w_{uv} = \exp\left(-\frac{\|\mathbf{X}_u^{t} - \mathbf{X}_v^{t}\|_2^2}{2\sigma^2}\right),
\end{equation}
where $\mathbf{X}_u^{t} \in \mathbb{R}^d$ represents the updated feature vector of node $u$ at iteration $t$, $\|\mathbf{X}_u^{t} - \mathbf{X}_v^{t}\|_2$ denotes the Euclidean distance between node features, and $\sigma \in \mathbb{R}^+$ is the Gaussian kernel bandwidth (learnable or fixed). The updated edge features are defined as:
\begin{equation}
	\mathbf{E}_{uv}^{t} = \mathbf{E}_{uv}^{\text{init}} \cdot w_{uv},
\end{equation}
where $\mathbf{E}_{uv}^{\text{init}} \in \mathbb{R}^{F_e}$ represents the initial edge feature vector.

\subsubsection*{Message Passing}
Node embeddings are updated via message passing:
\begin{equation}
	\hat{\mathbf{X}}_u^{t-1} = W^\top \mathbf{X}_u^{t-1} + \sum_{v \in \mathcal{N}(u)} \mathbf{X}_v^{t-1} \phi^e(E_{uv}^t),
\end{equation}
where $\hat{\mathbf{X}}_u^{t-1} \in \mathbb{R}^{d}$ is the intermediate feature of atom $u$ at iteration $t-1$, $W \in \mathbb{R}^{d \times d}$ is a learnable weight matrix, and $\mathcal{N}(u)$ is the set of neighboring nodes of atom $u$. 

Intermediate features are further updated using a gated recurrent unit (GRU):
\begin{equation}
	\mathbf{X}_u^{t} = \text{GRU}(\hat{\mathbf{X}}_u^{t-1}, H_u^{t-1}),
\end{equation}
where $\mathbf{X}_u^{t} \in \mathbb{R}^{d}$ is the updated feature of atom $u$ at iteration $t$, and $H_u^{t-1} \in \mathbb{R}^{d}$ denotes the previous hidden state of the GRU. After $T$ iterations, node features are concatenated with the original atomic features (residual connection):
\begin{equation}
	\mathbf{X} = \text{Concat}(\mathbf{X}^T, \mathbf{F_u}),
\end{equation}
where $\mathbf{X} \in \mathbb{R}^{N \times d'}$ and $d'$ denotes the concatenated feature dimension. The concatenated matrix is subsequently normalized.

\subsubsection*{Edge Weight Computation via Gaussian Kernel}
To construct the molecular graph, we compute the edge weight $w_{uv}$ between node (atom) $u$ and $v$ using a Gaussian kernel:
\begin{equation}
	\label{eq:updated_edge_weigth}
	w_{uv} = \exp\left(-\frac{\|\mathbf{X}_u - \mathbf{X}_v\|_2^2}{2\sigma^2}\right),
\end{equation}
where $\mathbf{X}_u, \mathbf{X}_v \in \mathbb{R}^d$ are the feature vectors of nodes $u$ and $v$, $\|\mathbf{X}_u - \mathbf{X}_v\|_2$ denotes their Euclidean distance, and $\sigma \in \mathbb{R}^+$ is the bandwidth parameter of the Gaussian kernel.

The resulting adjacency matrix of the molecular graph $\mathbf{W} \in \mathbb{R}^{N \times N}$ is defined as:
\begin{equation}
	\mathbf{W}[u, v] = 
	\begin{cases} 
		w_{uv}, & \text{if nodes } u \text{ and } v \text{ are connected by a chemical bond}, \\
		0, & \text{otherwise}.
	\end{cases}
\end{equation}

\subsection*{Community Detection via Leiden Algorithm}
Leiden algorithm is a graph partitioning method designed for community detection by maximizing modularity, aiming to identify densely connected substructures in molecular graphs.\cite{Traag2019} The modularity $Q$ is defined as:
\begin{equation}
	\label{modularity}
	Q = \frac{1}{2m} \sum_{u, v} \left[ w_{uv} - \gamma \frac{k_u k_v}{2m} \right] \delta(c_u, c_v),
\end{equation}
where $m = \sum_{u, v} w_{uv}$ is the total edge weight in the network, $k_u = \sum_{v} w_{uv}$ is the weighted degree of node $u$, $c_u$ is the community assignment of node $u$, and $\delta(c_u, c_v)$ is the Kronecker delta function, which equals 1 if $c_u = c_v$ and 0 otherwise. The parameter $\gamma$ controls the CG resolution: a larger $\gamma$ leads to more communities (i.e., CG beads).

Leiden algorithm improves upon Louvain method \cite{Blondel2008} by addressing its limitations such as disconnected communities and poor local minima. The key idea is to iteratively refine the community structure to ensure tightly connected intra-community nodes and sparsely connected inter-community nodes. The algorithm follows these steps:

1) Local moving phase: Each node is initially treated as its own community. Nodes are then moved between communities to locally maximize modularity.

2) Community aggregation: The resulting communities are contracted into super-nodes, forming a reduced graph.

3) Refinement: The process is repeated on the new graph until the modularity converges and no further improvement is possible.

Finally, the molecular graph is partitioned into $K$ communities:
\begin{equation}
	\{C_1, C_2, \dots, C_K\}, \quad \bigcup_{k=1}^{K} C_k = V, \quad C_i \cap C_j = \emptyset \quad \text{for } i \neq j.
\end{equation}

Each community $C_k$ corresponds to one coarse-grained group within the molecule.

\subsection*{Training}
\subsubsection*{Loss Function}
The total loss function for training the CG mapping model consists of two components: (1) triplet loss, and (2) predefined group pair loss. The overall loss is formulated as:
\begin{equation}
	\label{lossfunction}
	\mathcal{L} = \mathcal{L}_{\text{triplet}} +  \mathcal{L}_{\text{group}}.
\end{equation}
The triplet loss enforces feature proximity among atoms within the same CG group while maximizing separation between different groups. The predefined group pair loss ($\mathcal{L}_{\text{group}}$) preserves chemically meaningful groupings based on prior knowledge (e.g., SMARTS patterns \cite{Daylight_SMARTS}), ensuring that predefined groups are not inadvertently split during mapping. Together, these loss terms enable the model to automatically learn chemically reasonable CG mapping rules by balancing data-driven learning with prior knowledge constraints. The following sections detail the components of the loss function.

\subsubsection*{Triplet Loss}
The triplet loss ensures that atoms belonging to the same CG group are closer in the feature space, while atoms in different groups are farther apart:
\begin{equation}
	\mathcal{L}_{\text{triplet}} = \lambda_{\text{triplet}} \cdot \sum_{(i, j, k) \in \mathcal{T}} \left[ d(\mathbf{f}_i, \mathbf{f}_k) - d(\mathbf{f}_i, \mathbf{f}_j) + \text{margin} \right]_+,
\end{equation}
where $\lambda_{\text{triplet}}$ is a scaling factor, and $\mathcal{T}$ denotes the set of triplets $(i, j, k)$, with $i$ and $j$ belonging to the same CG group and $k$ to a different group. The feature vectors $\mathbf{f}_i, \mathbf{f}_j, \mathbf{f}_k \in \mathbb{R}^d$ represent atoms $i$, $j$, and $k$, respectively. The distance metric is defined as:
\begin{equation}
	d(\mathbf{f}_i, \mathbf{f}_j) = \exp\left(-\frac{\|\mathbf{f}_i - \mathbf{f}_j\|_2^2}{2\sigma^2}\right),
\end{equation}
where $\sigma$ is the Gaussian kernel bandwidth. The operator $[\cdot]_+$ denotes the ReLU function, ensuring non-negative loss values, and $\text{margin}$ is a hyperparameter controlling the minimum separation between intra- and inter-group distances.

\subsubsection*{Predefined Group Pair Loss}
The predefined group pair loss constrains atoms belonging to the same predefined chemical group (e.g., identified using SMARTS patterns \cite{Daylight_SMARTS}) to remain similarity in the feature space:
\begin{equation}
	\label{predefined_group_pair_loss}
	\mathcal{L}_{\text{group}} = \lambda_{\text{group}} \cdot \sum_{(i, j) \in \mathcal{P}} \left[ 1 - \exp\left(-\frac{\|\mathbf{f}_i - \mathbf{f}_j\|_2^2}{2\sigma^2}\right) \right],
\end{equation}
where $\lambda_{\text{group}}$ is a scaling factor, $\mathcal{P}$ denotes the set of atom pairs within the same predefined group, and $\sigma$ is the Gaussian kernel bandwidth.

\subsubsection*{Datasets}
ZINC is a publicly available molecular database comprising drug-like molecules and is widely used in cheminformatics, molecular generation, molecular graph learning, and virtual screening tasks. It provides a diverse set of molecules with varying structures and chemical properties, which supports robust generalization of the model \cite{Irwin2012}. In this work, we use the ZINC-250K subset \cite{Gomez-Bombarelli2018} containing approximately 250000 small molecules represented by their SMILES strings. Its structural diversity and chemical relevance make it well-suited for training and pretraining CG mapping models. 

Compared to the 744-molecule dataset used in DSGPM-TP, our finetuning dataset for MolCluster contains 488 molecular items. This reduction arises because MolCluster requires generation of 3D coordinates from atomic features, followed by construction of weighted molecular graphs using Gaussian kernels. However, in some cases, RDKit failed to converge to valid 3D conformations, preventing graph weight computation and thus reducing the dataset size. To ensure experimental consistency and reliability, only molecules with valid 3D structures were retained. All datasets are split into a ratio of 8:1:1 for training, validation, and testing, respectively.

\subsubsection*{Hyperparameter Optimization}
We employ random search to identify optimal hyperparameter settings based on performance on the validation set. Table~\ref{hyperparameters} summarizes the hyperparameters and their respective search ranges.

\begin{table}[h]
	\centering
	\caption{\enspace Hyperparameter descriptions and ranges.}
	\label{hyperparameters}
	\footnotesize
	\setlength{\tabcolsep}{4pt} 
	\renewcommand{\arraystretch}{1.2} 
		\begin{tabular}{lll}
			\hline
			\textbf{Name} & \textbf{Description} & \textbf{Range} \\
			\hline
			\texttt{batch\_size} & Input batch size & \{32, 64, 128\} \\
			\texttt{lr} & Initial learning rate & \{$10^{-4}$, $5 \times 10^{-4}$, $10^{-3}$\} \\
			\texttt{dropout} & Dropout ratio for the GNN & \{0, 0.1, 0.2, 0.3\} \\
			\texttt{n\_layer} & Number of hidden GNN layers & \{2, 4, 6, 8, 10, 12\} \\
			\texttt{n\_tlayer} & Number of hidden Transformer layers & \{1, 2, 3, 4, 5, 6\} \\
			\texttt{n\_head} & Number of heads in the multi-head Transformer & \{1, 2, 3, 4, 5, 6\} \\
			\texttt{hidden\_size} & Dimension of hidden GNN layers & \{64, 256\} \\
			$\lambda_\text{Loss}$ & Loss weight factor & \{0.01, 0.1, 1, 10, 100\} \\
			\texttt{init\_method} & Weight initialization method & \{None, Xavier, Kaiming\} \\
			\hline
		\end{tabular}
\end{table}

\backmatter

\section*{Supplementary information}

Further information about methods, group SMARTS and heat map of number of CG groups under different  resolution parameter $\gamma$.

\section*{Acknowledgements}

This work was supported by the National Natural Science Foundation of China under Grant No. 22422308.

\section*{Data Availability}

The source code and dataset can be obtained via GitHub (\url{https://github.com/JiangGroup/MolCluster}), allowing users to download, customize, and contribute to the program's development.




\bibliography{sn-bibliography.bib}
\bibliographystyle{sn}

\end{document}